\documentclass[11pt]{article}
\usepackage[margin=1in]{geometry}
\usepackage{booktabs}
\usepackage{amsmath}
\usepackage{graphicx}
\usepackage{xcolor}
\usepackage[hidelinks]{hyperref}
\usepackage{caption}
\usepackage{listings}

\newcommand{\repourl}{https://github.com/gauravapiscean/kernel-headroom}

\title{\bf How Much of a Real Workload Can LLM-Generated\\GPU Kernels Actually Reach?}
\author{
  Gaurav Agarwal\thanks{\texttt{gaurav.iit219@gmail.com}}
  \and
  Ashish Garg, PhD\thanks{\texttt{ashish.philly@gmail.com}}
  \and
  Isha Singhal\thanks{\texttt{ishasinghal22@gmail.com}}
}
\date{\today}

\begin{document}
\maketitle

\begin{abstract}
Language models can now write GPU kernels that outperform PyTorch. We evaluate
five model configurations on KernelBench level 1 and find that a frontier model
produces correct kernels for 91.1\% of problems and independently verified
speedups on 22 of 56, including three convolutions, with a median of
$1.235\times$. Open-weights models are far behind: the best reaches 30.4\%
correct with three verified speedups and solves zero convolutions.

We then ask a question the literature does not: what fraction of a real model's
wall clock do such kernels govern? Profiling seven workloads across three
domains, we find the \emph{addressable fraction} ranges from 8.9\% to 58.2\%. On
transformers, 80--86\% of runtime is spent in cuBLAS GEMM and FlashAttention,
bounding realistic end-to-end improvement at roughly 1\%, and the fraction
\emph{shrinks} with model scale. On recommenders it is 58.2\%, concentrated in a
single embedding kernel. We introduce DLRM-Bench, 12 recommender kernel problems
in KernelBench format, and measure a 41.7\% win rate at a $1.552\times$ median
there, projecting 8.63\% end-to-end.

Separately, we show that KernelBench's correctness check---\texttt{torch.allclose}
with an absolute tolerance---is satisfied by a tensor of zeros on 4 of 60 level-1
problems. Two kernels in our own results exploited this before we detected them,
including one scored at $283\times$ that wrote 0.3\% of its output buffer. We
propose scale-invariant replacements and release all 879 evaluations.
\end{abstract}

\section{Introduction}

Automated GPU kernel generation is an active area
\cite{concur,kernelscientist}, and KernelBench \cite{kernelbench} has become its
standard evaluation. Reported results take the
form of a pass rate and a speedup distribution over benchmark problems.

Those numbers answer ``how good are the kernels?'' They do not answer ``does it
matter?''. The quantity a practitioner needs is Amdahl's ratio \cite{amdahl},
\begin{equation}
\text{end-to-end gain} = \sum_{\text{ops}} s_i \left(1 - \frac{1}{\sigma_i}\right),
\label{eq:amdahl}
\end{equation}
where $s_i$ is operation $i$'s share of wall clock and $\sigma_i$ its speedup. A
$1.235\times$ speedup on an operation worth 3\% of runtime yields a 0.6\%
improvement; the same speedup on an operation worth 40\% yields 9\%. Benchmark
results supply $\sigma$ and are silent on $s$.

This paper measures $s$. Our contributions:

\begin{enumerate}
\item \textbf{The addressable fraction} (\S\ref{sec:amdahl}). Across seven
profiled workloads spanning transformers, CNNs and recommenders, it ranges from
8.9\% to 58.2\%---a $6.5\times$ spread---and is a property of the \emph{domain}
rather than the model or the method.
\item \textbf{A correctness exploit} (\S\ref{sec:exploit}) in a widely used
benchmark, with two worked examples that passed both its numerical check and its
anti-cheat check, and a proposed fix.
\item \textbf{A controlled backend measurement} (\S\ref{sec:backend}): holding
model, hardware, prompts and evaluation fixed, switching CUDA to Triton moves
pass@1 from 0.0\% to 10.7\%.
\item \textbf{DLRM-Bench} (\S\ref{sec:dlrm}), 12 recommender kernel problems, used
to verify that win rates transfer to a new domain rather than assuming they do.
\item \textbf{A taxonomy of ten measurement failures} (\S\ref{sec:bugs})
encountered while building the above, each of which produced a confident, wrong
conclusion before detection.
\end{enumerate}

\section{Experimental setup}

All experiments use two NVIDIA A100-80GB GPUs (Ampere, \texttt{sm\_80}), torch
2.14.0+cu130, CUDA 13.0 and triton 3.8.0, at bf16 precision throughout. The
benchmark is KernelBench level 1, problems 1--60 (single operations), with Triton \cite{triton}
as the primary target and CUDA C++ as a control.

Speedups are quoted against the \emph{fastest} of three baselines: eager
execution (cuBLAS), \texttt{torch.compile} \cite{pytorch2} with the inductor backend, and a
pre-recorded inductor timing set measured on the same hardware. Decoding is
greedy unless stated.

\paragraph{Denominator convention.} Four problems are excluded from all rates
because their correctness check is uninformative (\S\ref{sec:exploit}); all
percentages are over $n = 56$ unless noted.

\paragraph{Scale.} 879 evaluations and 874 generations across 19 experiments and
five model configurations: Qwen2.5-Coder-14B-Instruct, Qwen2.5-Coder-32B-Instruct,
DeepSeek-R1-Distill-Qwen-32B (MIT licensed), and GPT-5.6.

\section{Verification methodology}
\label{sec:verify}

Because the benchmark's own correctness check proved insufficient
(\S\ref{sec:exploit}), every claimed speedup was re-verified independently in an
isolated process against four criteria:

\begin{enumerate}
\item \textbf{Scale-invariant correctness}: relative Frobenius error
$\|\hat{y}-y\|_F / \|y\|_F < 2\times10^{-2}$ and maximum error normalised to
tensor magnitude $\max|\hat{y}-y| / \max|y| < 5\times10^{-2}$. Absolute
tolerances are unsafe when reference outputs are near zero.
\item \textbf{Fraction of output written}, compared against the reference's own
nonzero fraction. Catches kernels returning barely-touched buffers.
\item \textbf{Operation invariants} where they exist---softmax rows sum to one,
normalisations have unit norm. These cannot be satisfied by accident.
\item \textbf{Re-timing} with the benchmark's own timing function (cold L2, 5
warmup iterations, 100 trials, first discarded), against eager and
\texttt{torch.compile} measured in the same process.
\end{enumerate}

Two methodological requirements emerged from getting them wrong. Random-init
layers must be seeded immediately before constructing \emph{each} model, or
reference and candidate receive different weights; omitting this produced
apparent 140\% errors on three convolution kernels that were correct to bf16
rounding. And timing loops must not be hand-rolled: a custom loop that
synchronised and idled between trials allowed GPU clocks to drop and caused a
spurious retraction of a real result.

\section{Model comparison}
\label{sec:models}

\begin{table}[h]
\centering
\begin{tabular}{lrrr}
\toprule
Configuration & Correct & Faster & Convolutions \\
\midrule
Qwen2.5-Coder-14B, pass@1        & 10.7\% & 3.6\% & 0 / 8 \\
Qwen2.5-Coder-14B, 3 rounds      & 16.1\% & 5.4\% & 0 / 8 \\
Qwen2.5-Coder-32B, pass@1        & 17.9\% & 1.8\% & 0 / 8 \\
Qwen2.5-Coder-32B, 3 rounds      & 30.4\% & 5.4\% & 0 / 8 \\
Qwen2.5-Coder-32B, best-of-4     & 19.6\% & 5.4\% & 0 / 8 \\
DeepSeek-R1-Distill-Qwen-32B     &  7.1\% & 1.8\% & 0 / 8 \\
\textbf{GPT-5.6}                 & \textbf{91.1\%} & \textbf{39.3\%} & \textbf{7 / 8} \\
\bottomrule
\end{tabular}
\caption{KernelBench level 1, Triton, bf16, $n=56$. ``Faster'' is claimed before
independent verification.}
\end{table}

The gap between open and frontier models is not incremental. GPT-5.6's compile
failure rate is 0\% against 68\% for Qwen-32B. Convolutions are the sharpest
discriminator: no open model solved one in any experiment or round, while GPT-5.6
solved seven of eight.

Two secondary observations. Scaling 14B to 32B raised correctness but
\emph{lowered} the useful rate, the larger model writing correct but slower
kernels. And reasoning distillation at matched scale performed worse than code
specialisation (7.1\% versus 30.4\%).

\subsection{The backend effect}
\label{sec:backend}

Holding model, hardware, prompts, baselines and evaluation fixed and changing
only the target language moves pass@1 from 0.0\% (CUDA) to 10.7\% (Triton). CUDA
failures are 80\% compile errors in host-side scaffolding---C++ templates, pybind
naming, build configuration---so the model rarely reaches GPU logic at all.
Triton failures occur inside the kernel body, on indexing and masking. Reports of
a single KernelBench number should state which backend was used; the benchmark's
own prompt constructor defaults to Triton.

\subsection{Verified speedups}

Twenty-two kernels from GPT-5.6 passed full verification, with a median of
$1.235\times$ and a maximum of $3.057\times$. Three claimed speedups were
demoted on clean re-timing, indicating that single-shot harness timings run
optimistic. Where the mechanism is visible it is algorithmic: the
upper-triangular matmul writes exactly 50.01\% of its output, skipping the lower
triangle where PyTorch performs a dense matmul and masks.

\subsection{Interventions that did not work}

Asking the model to optimise an existing working kernel produced a portfolio
speedup of $1.000\times$ over 11 kernels and two rounds; a kernel running
$67\times$ slower than baseline was returned byte-identical. Exhaustive tile
search over 12 configurations for 9 kernels yielded $1.012\times$ aggregate with
zero kernels crossing from slower to faster---the model's stock configuration was
already optimal or within 2\% on seven of nine. Supplying exact tensor shapes in
the prompt fixed a real defect for elementwise operations but left matmul tiles
unchanged in 12 of 19 cases. Best-of-4 sampling tripled the useful rate but found
no new speedups.

\section{A correctness exploit}
\label{sec:exploit}

KernelBench establishes correctness with
\begin{lstlisting}
tolerance = get_tolerance_for_precision(precision)  # fp32: 1e-4, fp16/bf16: 1e-2
torch.allclose(output, output_new, atol=tolerance, rtol=tolerance)
\end{lstlisting}

This is an \emph{absolute} tolerance. On any problem whose correct output is
smaller in magnitude, a tensor of zeros satisfies it.

\begin{table}[h]
\centering
\begin{tabular}{lrcc}
\toprule
Problem & $\max|y|$ & fp32 (1e-4) & bf16 (1e-2) \\
\midrule
\texttt{23\_Softmax}         & 4.02e-06 & gameable & gameable \\
\texttt{37\_FrobeniusNorm}   & 3.98e-05 & gameable & gameable \\
\texttt{53\_Min\_reduction}  & 3.6e-03  & ok       & gameable \\
\texttt{39\_L2Norm}          & 6.8e-03  & ok       & gameable \\
\bottomrule
\end{tabular}
\caption{Four of 60 level-1 problems at bf16, two at fp32.}
\end{table}

\texttt{23\_Softmax} is $4096 \times 393216$, so every correct output element is
$\approx 2.5\times10^{-6}$---four thousand times under the bf16 tolerance. Two
kernels in our results exploited this:

\paragraph{Example A, scored $1.741\times$.} Applies softmax to 1024-element
chunks rather than rows. A correct softmax has rows summing to 1.0; this one's
sum to exactly $384.0 = 393216/1024$. Values remain under the tolerance, so
\texttt{allclose} passes. This result was treated as our best for two days.

\paragraph{Example B, scored $283\times$.} Row sums of 0.0026 with only 0.3\% of
the output buffer written. Investigated only because $283\times$ is physically
impossible for a memory-bound operation.

Neither triggered the static anti-cheat checker, which looks for library calls;
these kernels call none and simply do not perform the work. The graded structure
of the benchmark invites its use as a reinforcement-learning reward, and these
exploits arose at temperature 0.8 with no optimisation pressure. \S\ref{sec:verify}
gives the checks we propose instead; any one of the three detects both examples.

\section{The addressable fraction}
\label{sec:amdahl}

We profiled real workloads and classified every CUDA kernel as
\textsc{already-optimal} (cuBLAS GEMM, FlashAttention \cite{flashattention}, cuDNN convolution),
\textsc{addressable} (elementwise, normalisation, softmax, reductions,
embedding), \textsc{overhead} (copies, layout transforms) or unclassified.

\begin{table}[h]
\centering
\begin{tabular}{llrrr}
\toprule
Workload & Mode & Already optimal & \textbf{Addressable} & Realistic \\
\midrule
DeepSeek-R1-Distill-Llama-70B & inference & 85.9\% & \textbf{8.9\%}  & 1.32\% \\
Qwen2.5-Coder-14B             & training  & 64.4\% & 16.1\% & 2.39\% \\
Qwen2.5-Coder-14B             & inference & 79.8\% & 16.8\% & 2.49\% \\
DLRM                          & training  & 32.5\% & \textbf{41.9\%} & 6.21\% \\
ResNet-50                     & inference & 36.8\% & \textbf{46.1\%} & --- \\
ResNet-50                     & training  & 29.0\% & \textbf{55.3\%} & --- \\
\textbf{DLRM}                 & inference & 37.4\% & \textbf{58.2\%} & \textbf{8.63\%} \\
\bottomrule
\end{tabular}
\caption{A $6.5\times$ spread using identical kernels and methods. ``Realistic''
applies the measured in-domain win rate and median speedup via
Eq.~\ref{eq:amdahl}.}
\label{tab:amdahl}
\end{table}

\paragraph{Transformers.} Every configuration lands between 8.9\% and 16.8\%
addressable. The fraction \emph{shrinks with scale}: at 70B the GEMMs are larger
and more compute-bound, and the top four cuBLAS kernels alone account for 82.8\%
of wall clock. This is adverse for any commercial case, since the workloads with
the largest compute budgets have the least addressable work.

Training is not better. Its addressable fraction is 16.1\% against inference's
16.8\%, and a further 18.2\% is consumed by the optimizer step, which is
unbeatable by construction:
$14.7\times10^9$ parameters $\times$ 6 bytes (read parameter, read gradient,
write parameter) $= 88$\,GB; at $\approx 1.5$\,TB/s achievable bandwidth this
predicts 59\,ms against 56\,ms measured.

\paragraph{CNNs and recommenders.} The addressable fraction is
$2.7$--$3.5\times$ larger, and concentrated rather than scattered. BatchNorm
alone is 44.5\% of a CNN training step; \texttt{EmbeddingBag\_updateOutputKernel\_sum}
alone is 37.4\% of a DLRM inference step, and embedding plus indexing operations
total 49.5\% of DLRM training. A concentrated op family is a substantially better
engineering target than a long tail.

\section{DLRM-Bench}
\label{sec:dlrm}

Table~\ref{tab:amdahl}'s recommender projections initially used a win rate and
median speedup transferred from KernelBench---an assumption about a domain for
which no kernel had been generated. We therefore built DLRM-Bench, targeting the kernel surface of a DLRM-style
recommender \cite{dlrm}: 12
KernelBench-format problems covering EmbeddingBag sum and mean pooling, plain
gather, multi-table lookup, sparse scatter-add, gather with LayerNorm, weighted
pooling, pairwise dot interaction, row normalisation, SparseLengthsSum, top-$k$
ranking, and gather with bias and ReLU.

\begin{table}[h]
\centering
\begin{tabular}{lrr}
\toprule
 & Transferred assumption & Measured in-domain \\
\midrule
Win rate       & 37.0\% & \textbf{41.7\%} (5 of 12) \\
Median speedup & $1.235\times$ & $\mathbf{1.552\times}$ \\
Maximum        & $3.057\times$ & $\mathbf{6.917\times}$ \\
\bottomrule
\end{tabular}
\caption{The assumption held and was conservative.}
\end{table}

The largest win, $6.917\times$ on multi-table lookup, verifies bit-exact with
100\% of output written. Its mechanism is visible in the generated source: one
kernel performs all eight table lookups and writes the concatenated output
directly, replacing eight separate gathers and a concatenation. This fusion
opportunity exists in recommenders and is structurally absent from transformers.

The four losses are equally informative. \texttt{torch.compile} already beats
eager by $4\times$ on embedding lookups ($0.182 \to 0.044$\,ms), so the bar in
this domain is considerably higher than naive PyTorch.

\paragraph{Dogfooding.} Our first draft initialised embedding tables such that
one problem produced a maximum output of 0.0125---only $1.25\times$ above the
bf16 tolerance, reproducing the exact defect of \S\ref{sec:exploit}. After
correcting the initialisation scale, zero of 12 problems are gameable, with a
minimum margin of $38.9\times$.

\section{Measurement failures}
\label{sec:bugs}

Ten measurement errors were identified over 19 experiments. Each produced a
confident, wrong conclusion before detection. We report them because the rate is
itself a finding about this kind of work.

Representative examples: running each backend's validator manually reported
``100\% cheated'' when nothing had been tested; omitting the \texttt{backend}
argument to the evaluator would have reported approximately 100\% Triton compile
failure when the harness could not execute \emph{any} Triton kernel; flooring
relative error at $10^{-12}$ made three correct convolutions appear wrong by a
factor of $10^{11}$; and omitting a seed before constructing the candidate model
gave it different random convolution weights than the reference.

\textbf{Errors in this domain are symmetric.} Two of the ten were \emph{hiding}
real results rather than manufacturing false ones, and both had to be fixed before
three genuine convolution speedups became visible. A positive control---a
hand-written known-good kernel pushed through the full pipeline---detected four
of the ten.

\section{Related work}

KernelBench \cite{kernelbench} provides the benchmark this work builds on, and a
growing body of systems target it \cite{concur,kernelscientist}. The
correctness-performance gap is established elsewhere and we do not claim it.
\emph{Correct but Slow} \cite{correctslow} studies 22 Triton and TileLang kernels
and reports a TileLang LayerNorm passing KernelBench's correctness check while
running $300\times$ slower than PyTorch. \emph{KernelBenchX} \cite{kernelbenchx}
evaluates 176 tasks across 15 categories, finding 46.6\% of correct kernels slower
than eager, that iterative refinement raises compile rate while \emph{lowering}
average speedup, and that task category explains roughly three times more variance
in correctness than method choice---a within-benchmark analogue of the domain
effect in \S\ref{sec:amdahl}.

Our \S\ref{sec:models} results are consistent with these and add little. What this
work adds is the denominator: prior studies measure how good generated kernels
are, while none measures what fraction of a real model's runtime those kernels can
touch. \S\ref{sec:exploit} also reports a distinct failure mode---kernels that are
\emph{incorrect and scored correct}, rather than correct and slow.

\section{Limitations}

One host and one GPU type. KernelBench level 1 only; levels 2 and 3 are closer to
production code. One batch size and shape per profile, and the compute/bandwidth
balance shifts with both. One model per family. The CNN is a ResNet-50 topology
written in plain \texttt{torch.nn} rather than a deployed vision pipeline. No
training was performed; distillation was scoped but not run. The frontier model's
91.1\% will age, though the structural result should not. Bucket classification is
pattern-based on kernel names and therefore approximate; one misclassification was
found and corrected, and 99.5\% or more of measured time was classified in every
profile.

\section{Conclusion}

Frontier models write competitive GPU kernels today, and open models of the sizes
we tested do not come close. But benchmark performance does not transfer to
end-to-end impact on transformers, where the addressable fraction is 8.9--16.8\%
and shrinks with scale. It does transfer on recommenders, where 58.2\% is
addressable and a purpose-built benchmark confirms a 41.7\% win rate.

The practical implication is to profile a domain's addressable fraction
\emph{before} generating kernels for it. Transformers are the worst available
target precisely because they are the best optimised.

\paragraph{Artifacts.} All 879 evaluations and 874 generations, the independent
verifier, the profilers and DLRM-Bench are available at \url{\repourl}.

\end{document}